# *J*-matrix method of scattering in one dimension: The relativistic theory

A. D. Alhaidari

*Saudi Center for Theoretical Physics, P.O. Box 32741, Jeddah 21438, Saudi Arabia*

**Abstract**: We make a relativistic extension of the one-dimensional *J*-matrix method of scattering. The relativistic potential matrix is a combination of vector, scalar, and pseudo-scalar components. These are non-singular short-range potential functions (not necessarily analytic) such that they are well represented by their matrix elements in a finite subset of a square integrable basis set that supports a tridiagonal symmetric matrix representation for the free Dirac operator. Transmission and reflection coefficients are calculated for different potential coupling modes. This is the first of a two- paper sequence where we develop the theory in this part then follow it with applications in the second.

***This work is dedicated to the memory of my friend, colleague and collaborator the late Mohammed S. Abdelmonem.***

**PACS**: 03.65.Nk, 03.65.Pm, 11.80.-m, 03.65.Fd

**Keywords**: *J*-matrix, relativistic scattering, (1+1) space-time, Dirac equation, vector, scalar, pseudo-scalar.

## I. INTRODUCTION

Not long ago, we presented a formulation of the nonrelativistic scattering problem in one dimension based on the *J*-matrix method [1]. We found that the usual thought of simplicity of the 1D scattering problem, when compared to 3D, is not manifest as indicated by the nontrivial and highly rich structure that emerged. For example, there will always be two scattering channels that are non-trivially coupled. The two channels decouple only if the interaction potential has a definite parity (i.e., being either odd or even function in the configuration space coordinate *x*). In that formulation, we deployed the analytic power built in the *J*-matrix method where the theory of orthogonal polynomials plays an important role [2-8]. Additionally, we exploited the efficiency of numerical schemes associated with tridiagonal matrices that resulted in a convergent and accurate calculation of the reflection and transmission amplitudes. These schemes include Gauss quadrature and continued fraction techniques [9-13]. We have also demonstrated that the formulation could be used as an alternative means for resolving issues related to bound states, resonances and scattering phenomena in one-dimensional systems. Moreover, we have shown that it constitutes a viable alternative to the classical treatment of the 1D scattering problems and that it could also help unveil new and interesting applications. We believe that the relativistic extension of that theory will have similar qualities. Therefore, the objective here is to present a relativistic extension of the work in [1]. That is, we would like to formulate the relativistic *J*-matrix method of scattering in 1+1 space-time dimension. This theory becomes relevant when studying one-dimensional models at higher energies compared to the rest mass energy or at strong coupling. Moreover, it offers various choices of coupling modes for the same potential configuration. These modes include vector, scalar, pseudo-scalar, and a mix of these. The reader is strongly advised to consult Ref. [1] for background, introduction and notation.

Consequently, we are interested here in the scattering solution of the following time-independent Dirac equation in 1+1 Minkowski space-time

$$(H_0 + \mathcal{V})|\psi\rangle = \varepsilon|\psi\rangle, \tag{1}$$

where $\psi$ is a two-component wavefunction and $\varepsilon$ is the relativistic energy. $H_0$ is the free Dirac operator and $\mathcal{V}$ is a 2×2 potential matrix. In the relativistic units $\hbar = c = 1$ and in a typical representation of the Dirac gamma matrices in 1+1 space=time, where we can take $\gamma_0 = \sigma_3 = \begin{pmatrix} 1 & 0 \\ 0 & -1 \end{pmatrix}$ and $\gamma_1 = i\sigma_1 = i\begin{pmatrix} 0 & 1 \\ 1 & 0 \end{pmatrix}$, we can write[1]

$$H_0 = \begin{pmatrix} M & -\frac{d}{dx} \\ \frac{d}{dx} & -M \end{pmatrix}, \text{ and } \mathcal{V}(x) = \begin{pmatrix} V+S & U+iQ \\ U-iQ & V-S \end{pmatrix}, \tag{2}$$

where $M$ is the rest mass of the Dirac particle. The two-vector potential is $\vec{A}(x) = (A_0, A_1) = (V, Q)$, the scalar potential is $S(x)$ and the pseudo-scalar is $U(x)$. All these potentials are assumed to be non-singular and of short range such that they become vanishingly small for all $|x| > X$, where $X$ is greater than or equal to the range of the potential. We specialize to the case where the space component of the vector potential is gauged away (i.e., $Q = 0$) and we consider positive energy scattering, where $\varepsilon > M$. Negative energy scattering is for $\varepsilon < -M$, whereas bound states correspond to $|\varepsilon| \leq M$.

## II. *J*-MATRIX SOLUTION OF THE REFERENCE PROBLEM

The reference problem is defined by the solution of the free Dirac equation $H_0|\chi\rangle = \varepsilon|\chi\rangle$, which reads

$$\begin{pmatrix} M-\varepsilon & -\frac{d}{dx} \\ \frac{d}{dx} & -M-\varepsilon \end{pmatrix} \begin{pmatrix} \chi^\uparrow \\ \chi^\downarrow \end{pmatrix} = 0. \tag{3}$$

This equation results in the following relation between the upper ($\chi^\uparrow$) and lower ($\chi^\downarrow$) components of the reference wavefunction

$$\chi^{\downarrow\uparrow} = \frac{1}{M \pm \varepsilon} \frac{d\chi^{\uparrow\downarrow}}{dx}, \tag{4}$$

---

[1] Multiplying one of the two spinor components by an i results in an equivalent representation of the Dirac Hamiltonian whose matrix elements are identical to those in Eq. (2) except that the off-diagonal elements $\{\mp\frac{d}{dx}, U \pm iQ\}$ are replaced by $\{i\frac{d}{dx}, Q \mp iU\}$.

which is not valid for $\varepsilon = \mp M$, respectively. In this equation, the left (right) vertical arrow goes with the top (bottom) sign, respectively. Substituting (4) back in Eq. (3), gives the following second order differential equation

$$\left(\frac{d^2}{dx^2} + \varepsilon^2 - M^2\right)\chi^{\uparrow\downarrow} = 0.$$ (5)

Since $\varepsilon = +M$ ($\varepsilon = -M$) belongs to the positive (negative) energy spectrum, then the positive energy subspace of interest to this work corresponds to the top signs and left vertical arrows. Consequently, we obtain the following two independent positive energy solutions for the reference Dirac equation (3)

$$S^+(x) = \mathcal{A}\begin{pmatrix} \cos kx \\ -\omega \sin kx \end{pmatrix},$$ (6a)

$$S^-(x) = \mathcal{B}\begin{pmatrix} \sin kx \\ \omega \cos kx \end{pmatrix},$$ (6b)

where $k = \sqrt{\varepsilon^2 - M^2}$, $\omega = \sqrt{\frac{\varepsilon - M}{\varepsilon + M}}$, $\mathcal{A}$ and $\mathcal{B}$ are arbitrary real constants. In the terminology of the *J*-matrix method, we refer to these as the sine-like solutions. However, as in the non-relativistic 1D *J*-matrix [1], we end up with two scattering channels: an even channel and an odd channel to which $S^+(x)$ and $S^-(x)$ belong, respectively. Due to the finite range of the potential, the boundary conditions for the scattering problem, which is defined by the positive energy solution of Eq. (1) at infinity where $\mathcal{V} = 0$, are

$$\lim_{x \to -\infty} \psi(x,\varepsilon) = \begin{pmatrix} 1 \\ i\omega \end{pmatrix} e^{ikx} + R\begin{pmatrix} 1 \\ -i\omega \end{pmatrix} e^{-ikx},$$ (7a)

$$\lim_{x \to +\infty} \psi(x,\varepsilon) = T\begin{pmatrix} 1 \\ i\omega \end{pmatrix} e^{ikx}.$$ (7b)

These represent a normalized flux of relativistic particles with energy $\varepsilon$ incident on the potential region from left that gets partially reflected with an amplitude $R(\varepsilon)$ and partially transmitted with an amplitude $T(\varepsilon)$. These amplitudes depend not only on the energy but also on the potential parameters. Unitarity of the problem results in the current (particle flux) conservation equation, $|T|^2 + |R|^2 = 1$.

Following the standard formulation of the *J*-matrix method [2-8], we expand the reference solution (6) in a complete spinor basis for the two channels, $\{\phi_n^\pm\}_{n=0}^\infty$. These bases should be complete in the domain of $H_0$ and are required to carry a tridiagonal matrix representation for the reference wave operator, $J = H_0 - \varepsilon$. Now, we write

$$S^\pm(x) = \sum_{n=0}^\infty s_n^\pm(\varepsilon)\phi_n^\pm(x),$$ (8)

where $\{s_n^\pm\}$ are the expansion coefficients. The lower component of the spinor basis, $\phi_n^{\pm\downarrow}$, must be related to the upper, $\phi_n^{\pm\uparrow}$, by relation (4) with the top sign. This makes the lower component of the basis energy dependent. Following the same scheme as in the non-relativistic case [1] (however, here for two components instead of one) we obtain the following two-component spinor basis for the even channel

$$\phi_n^+(x,\varepsilon) = A_n e^{-y^2/2} \begin{pmatrix} H_{2n}(y) \\ \frac{\lambda}{M+\varepsilon}\left[2nH_{2n-1}(y) - \tfrac{1}{2}H_{2n+1}(y)\right] \end{pmatrix}, \qquad (9)$$

where $y = \lambda x$, $\lambda$ is a real positive scale parameter of inverse length dimension, $H_n(y)$ is the Hermite polynomial of order $n$, and the normalization constants is $A_n = \left[\pi^{1/4} 2^n \sqrt{(2n)!}\right]^{-1}$. Thus, we can write

$$\phi_n^{+\downarrow}(x) = \frac{\lambda}{M+\varepsilon}\left[\sqrt{n}\,\phi_{n-\frac{1}{2}}^{+\uparrow}(x) - \sqrt{n+\tfrac{1}{2}}\,\phi_{n+\frac{1}{2}}^{+\uparrow}(x)\right]. \qquad (10)$$

On the other hand, the spinor basis for the odd channel is obtained from that of the even channel (9) as follows

$$\phi_n^-(x,\varepsilon) = \phi_{n+\frac{1}{2}}^+(x,\varepsilon). \qquad (11)$$

Thus, $\langle \phi_n^+ | \phi_m^- \rangle = \langle \phi_n^- | \phi_m^+ \rangle = 0$ for all integers $n$ and $m$. Moreover, using the orthogonality relation of the Hermite polynomials [14,15] and the integral formula in [16], we obtain the following energy-dependent expansion coefficients for the even and odd channels

$$s_n^+(\varepsilon) = (-1)^n \sqrt{2\pi}\,\mathcal{A}\,A_n e^{-\mu^2/2} H_{2n}(\mu), \qquad (12a)$$

$$s_n^-(\varepsilon) = (-1)^n \sqrt{2\pi}\,\mathcal{B}\,A_{n+\frac{1}{2}} e^{-\mu^2/2} H_{2n+1}(\mu), \qquad (12b)$$

where $\mu = k/\lambda$ and we took the configuration space integral measure as $\lambda\int_{-\infty}^{+\infty} dx$ (refer to Sec. 2 of [1] for details). These two coefficients are related as $s_n^-(\varepsilon) = -i\frac{\mathcal{B}}{\mathcal{A}} s_{n+\frac{1}{2}}^+(\varepsilon)$ and they satisfy the following differential equation in the energy

$$\left(\frac{d^2}{d\mu^2} - \mu^2 + 4n + 2 \mp 1\right) s_n^\pm(\varepsilon) = 0. \qquad (13)$$

Using the recursion relation, differential property, and orthogonality of the Hermite polynomials [14,15], we calculate the tridiagonal matrix representation of the reference wave operator in the even subspace as

$$J^+_{n,m}(\varepsilon) \equiv \langle \phi^+_n | (H_0 - \varepsilon) | \phi^+_m \rangle$$
$$= \frac{\lambda^2}{\varepsilon + M} \left\{ \left[ (2n + \tfrac{1}{2}) - \mu^2 \right] \delta_{n,m} - \sqrt{n(n - \tfrac{1}{2})}\, \delta_{n,m+1} - \sqrt{(n+1)(n + \tfrac{1}{2})}\, \delta_{n,m-1} \right\} \quad (14a)$$

On the other hand, the matrix elements in the odd subspace, $J^-_{n,m} \equiv \langle \phi^-_n | J | \phi^-_m \rangle$, is obtained from those in (14a) by the replacement $n \to n + \tfrac{1}{2}$ and $m \to m + \tfrac{1}{2}$. That is,

$$J^-_{n,m}(\varepsilon) \equiv \langle \phi^-_n | (H_0 - \varepsilon) | \phi^-_m \rangle$$
$$= \frac{\lambda^2}{\varepsilon + M} \left\{ \left[ (2n + \tfrac{3}{2}) - \mu^2 \right] \delta_{n,m} - \sqrt{n(n + \tfrac{1}{2})}\, \delta_{n,m+1} - \sqrt{(n+1)(n + \tfrac{3}{2})}\, \delta_{n,m-1} \right\} \quad (14b)$$

Since the reference wave operator is an even matrix function[2], then $\langle \phi^+_n | J | \phi^-_m \rangle = \langle \phi^-_n | J | \phi^+_m \rangle = 0$. Moreover, as consequence of (14) the reference wave equation $J|\chi\rangle = 0$ becomes equivalent to the algebraic equation $\sum_m J^\pm_{nm} s^\pm_m = 0$, which results in the following symmetric three-term recursion relation for $s^-_n(\varepsilon)$

$$\mu^2 s^+_n = (2n + \tfrac{1}{2}) s^+_n - \sqrt{n(n - \tfrac{1}{2})}\, s^+_{n-1} - \sqrt{(n+1)(n + \tfrac{1}{2})}\, s^+_{n+1}, \quad n = 1, 2, 3, \ldots, \quad (15a)$$

$$\mu^2 s^+_0 = \tfrac{1}{2} s^+_0 - \sqrt{\tfrac{1}{2}}\, s^+_1. \quad (15b)$$

The recurrence relation for $s^-_n(\varepsilon)$ is obtained from (15) by the replacement $n \to n + \tfrac{1}{2}$. That is,

$$\mu^2 s^-_n = (2n + \tfrac{3}{2}) s^-_n - \sqrt{n(n + \tfrac{1}{2})}\, s^-_{n-1} - \sqrt{(n+1)(n + \tfrac{3}{2})}\, s^-_{n+1}, \quad n = 1, 2, 3, \ldots, \quad (16a)$$

$$\mu^2 s^-_0 = \tfrac{3}{2} s^-_0 - \sqrt{\tfrac{3}{2}}\, s^-_1. \quad (16b)$$

Now, the scattering information for the even and odd channels is contained in the difference at infinity between the phases of $S^\pm(x)$ and those of the complementary solutions $C^\pm(x)$, which will be obtained next. This phase difference depends on the energy and potential parameters. In the absence of interaction ($\mathcal{V} = 0$) it is $\pi/2$ for all energies. The complementary solutions (known in the *J*-matrix language as the cosine-like solutions) are written as $C^\pm(x) = \sum_n c^\pm_n(\varepsilon) \phi^\pm_n(x)$, where the coefficients $c^\pm_n(\varepsilon)$ are required to satisfy the following conditions:

(1) Be a second independent solution of the energy differential equation (13).

---

[2] An even 2×2 matrix function $Q(x)$ must have its two diagonal elements $Q_{11}(x)$ and $Q_{22}(x)$ as even functions of $x$ and its two off-diagonal elements $Q_{12}(x)$ and $Q_{21}(x)$ as odd functions of $x$. On the other hand, the elements of an odd 2×2 matrix function must have the opposite respective parity.

(2) Be a second independent solution of the three-term recursion relation (15a) and (16a) with initial relations that differ from (15b) and (16b).

(3) The initial relation for the recursion of $c_n^\pm(\varepsilon)$ is chosen such that it makes the complementary reference wavefunctions $C^\pm(x)$ asymptotically sinusoidal and identical to $S^\pm(x)$ but with a phase difference of $\frac{\pi}{2}$.

These conditions result in the following expansion coefficients for $C^\pm(x)$:

$$c_n^+(\varepsilon) = 2\mathcal{A}\sqrt{2\Gamma(n+1)/\Gamma(n+\tfrac{1}{2})}\,\mu e^{-\mu^2/2}\,{}_1F_1\left(-n+\tfrac{1}{2};\tfrac{3}{2};\mu^2\right), \tag{17a}$$

$$c_n^-(\varepsilon) = \mathcal{B}\sqrt{2\Gamma(n+1)/\Gamma(n+\tfrac{3}{2})}\,e^{-\mu^2/2}\,{}_1F_1\left(-n-\tfrac{1}{2};\tfrac{1}{2};\mu^2\right), \tag{17b}$$

where ${}_1F_1(a;c;z)$ is the confluent hypergeometric function. For the details of these calculations, one may consult Ref. [1]. The contiguous relation of ${}_1F_1(a;c;z)$ that reads [17]

$$z\,{}_1F_1(a;c;z) = (c-2a)\,{}_1F_1(a;c;z) + (a-c)\,{}_1F_1(a-1;c;z) + a\,{}_1F_1(a+1;c;z), \tag{18}$$

could be used to verify that (17a) and (17b) satisfy the three-term recursion relation (15a) and (16a) for $n = 1, 2, 3, \ldots$, respectively. On the other hand, the initial relations that replace (15b) and (16b) are respectively as follows

$$\mu^2 c_0^+ = \tfrac{1}{2} c_0^+ - \sqrt{\tfrac{1}{2}}\, c_1^+ + \frac{\sqrt{2}\mathcal{A}}{\pi^{1/4}}\,\mu e^{\mu^2/2}. \tag{19}$$

$$\mu^2 c_0^- = \tfrac{3}{2} c_0^- - \sqrt{\tfrac{3}{2}}\, c_1^- - \frac{\mathcal{B}}{\pi^{1/4}}\,e^{\mu^2/2}. \tag{20}$$

A simple argument in support of these findings is the observation that, in one of its representations in terms of the confluent hypergeometric function, the Hermite polynomial could be written as follows

$$H_n(x) = 2^n\sqrt{\pi}\left[\frac{1}{\Gamma\left(\frac{1-n}{2}\right)}\,{}_1F_1\left(-\tfrac{n}{2};\tfrac{1}{2};x^2\right) - \frac{2x}{\Gamma\left(-\tfrac{n}{2}\right)}\,{}_1F_1\left(\tfrac{1-n}{2};\tfrac{3}{2};x^2\right)\right]. \tag{21}$$

Therefore, for even degrees this relation gives $(-1)^n A_n e^{-\mu^2/2} H_{2n}(\mu) \propto s_n^+(\varepsilon) + \frac{(-1)^n}{\Gamma(-n)\Gamma(n+1)} c_n^+(\varepsilon)$ in which case only the $s_n^+(\varepsilon)$ term survives since $\Gamma(-n)$ blows up. Similarly, for odd degrees the same relation gives $(-1)^n A_{n+1/2} e^{-\mu^2/2} H_{2n+1}(\mu) \propto s_n^-(\varepsilon) - \frac{(-1)^n}{\Gamma(-n)\Gamma(n+1)} c_n^-(\varepsilon)$ in which case only the $s_n^-(\varepsilon)$ survives. Now, the cosine-like solutions have the following useful asymptotic properties

$$\lim_{x\to\pm\infty} \tfrac{1}{\mathcal{A}} C^+(x) = \pm \begin{pmatrix} \sin kx \\ \omega \cos kx \end{pmatrix} = \pm \tfrac{1}{\mathcal{B}} S^-(x), \tag{22a}$$

$$\lim_{x\to\pm\infty} \tfrac{1}{\mathcal{B}} C^-(x) = \pm \begin{pmatrix} \cos kx \\ -\omega \sin kx \end{pmatrix} = \pm \tfrac{1}{\mathcal{A}} S^+(x). \tag{22b}$$

We have shown elsewhere [18] that if we eliminate the lowest terms in the series and construct the truncated series $S_N^\pm(x) \equiv \sum_N^\infty s_n^\pm(\varepsilon)\phi_n^\pm(x,\varepsilon)$ and $C_N^\pm(x) \equiv \sum_N^\infty c_n^\pm(\varepsilon)\phi_n^\pm(x,\varepsilon)$, for some large enough integer $N$, then we obtain vanishingly small values for these functions within a finite region symmetric about the origin. The size of this region increases with the number of eliminated terms, $N$ [18]. In fact, there is a perfect parallel between the limit as $N \to \infty$ of these functions and the limit as $|x| \to \infty$, which we can exploit to write [1]

$$\lim_{N\to\infty} \left\{ \tfrac{1}{2}\left[\tfrac{1}{\mathcal{A}}S_N^+(x) \pm \tfrac{1}{\mathcal{B}}C_N^-(x)\right] + \tfrac{i}{2}\left[\tfrac{1}{\mathcal{B}}S_N^-(x) \pm \tfrac{1}{\mathcal{A}}C_N^+(x)\right] \right\} = \begin{cases} \begin{pmatrix} 1 \\ i\omega \end{pmatrix} e^{ikx} & , x \to \pm\infty \\ 0 & , x \to \mp\infty \end{cases} \tag{23}$$

As we shall see below, this property will proof very valuable in the construction of the complete solution of the problem such that it satisfies the asymptotic boundary conditions (7a) and (7b). Next, we will augment the kinematics obtained above by the dynamics coming from the interaction of the Dirac particle with the short range potential matrix $\mathcal{V}(x)$ and identify the scattering amplitudes $R(\varepsilon)$ and $T(\varepsilon)$.

### III. INTERACTION AND THE SCATTERING AMPLITUDES

The potential matrix $\mathcal{V}(x)$ is short-range such that it is well represented by its matrix elements in a large enough subset of a square integrable spinor basis set, $\{\xi_n(x)\}$. This set must be complete and should span the solution space of the original Dirac equation (1). Moreover, it is required to sample the potential in the interior scattering region $\left(|x| < X\right)$ to the desired degree of accuracy. We write the *complete* spinor wavefunction, $\psi(x,\varepsilon)$, as the following infinite bounded series in this complete basis set

$$\psi(x,\varepsilon) = \sum_{m=-\infty}^{+\infty} a_m(\varepsilon)\xi_m(x); \quad \sum_{m=-\infty}^{+\infty} |\xi_m\rangle\langle\xi_m| = 1. \tag{24}$$

Due to the finite range of the scattering potential, we split configuration space into three regions: the middle region $\left(|x| \leq X\right)$, left region $\left(x < -X\right)$, and right region $\left(x > X\right)$. In analogy to this splitting and in a similar manner to the nonrelativistic theory [1], we also split the algebraic function space spanned by $\{\xi_n(x)\}_{n=-\infty}^{\infty}$ into middle ($-N \leq n < N$), left ($n < -N$), and right ($n \geq N$) subspaces. Therefore, we rewrite the total wavefunction (24) in these three subspaces as follows

$$\psi(x,\varepsilon) = \sum_{m=-\infty}^{-N-1} b_{-m-1}^{-}(\varepsilon)\xi_m(x) + \sum_{m=-N}^{N-1} a_m(\varepsilon)\xi_m(x) + \sum_{m=N}^{+\infty} b_m^{+}(\varepsilon)\xi_m(x). \tag{25}$$

The size of the middle subspace, whose dimension is 2N, should be large enough to give an accurate matrix representation of the short-range scattering potential. Moreover, it should be large enough to give a faithful representation for the asymptotic sinusoidal limit of the total wavefunction as $|x| \to \infty$. Using the asymptotic properties of the sine-like and cosine-like solutions given by Eq. (23), we can write the boundary conditions (7a) and (7b) as

$$\lim_{x \to +\infty} \psi(x,\varepsilon) = T\binom{1}{i\omega}e^{ikx} \approx T\sum_{n=N}^{\infty}\left[\tfrac{1}{2\mathcal{A}}\left(s_n^{+} + ic_n^{+}\right)\phi_n^{+} + \tfrac{1}{2\mathcal{B}}\left(c_n^{-} + is_n^{-}\right)\phi_n^{-}\right], \text{ and} \tag{7a'}$$

$$\lim_{x \to -\infty} \psi(x,\varepsilon) = \binom{1}{i\omega}e^{ikx} + R\binom{1}{-i\omega}e^{-ikx} \approx \sum_{n=N}^{\infty}\left[\tfrac{1}{2\mathcal{A}}\left(s_n^{+} - ic_n^{+}\right)\phi_n^{+} - \tfrac{1}{2\mathcal{B}}\left(c_n^{-} - is_n^{-}\right)\phi_n^{-}\right]$$
$$+ R\sum_{n=N}^{\infty}\left[\tfrac{1}{2\mathcal{A}}\left(s_n^{+} + ic_n^{+}\right)\phi_n^{+} - \tfrac{1}{2\mathcal{B}}\left(c_n^{-} + is_n^{-}\right)\phi_n^{-}\right] \tag{7b'}$$

for large enough N. Comparing these asymptotic solutions represented as sums with the corresponding sum in Eq. (25) we deduce the following

$$\xi_m(x) = \begin{cases} \phi_m^{+}(x) &, m \geq N \\ \phi_{-m-1}^{-}(x) &, m < -N \end{cases} \tag{26}$$

$$b_m^{+}(\varepsilon) = W_{+}(\varepsilon)p_m^{+}(\varepsilon) + p_m^{-}(\varepsilon), \tag{27a}$$

$$b_m^{-}(\varepsilon) = W_{-}(\varepsilon)q_m^{+}(\varepsilon) - q_m^{-}(\varepsilon), \tag{27b}$$

where $W_{\pm}(\varepsilon) = T \pm R$, $p_m^{\pm}(\varepsilon) = \tfrac{1}{2\mathcal{A}}\left(s_m^{+} \pm ic_m^{+}\right)$, and $q_m^{\pm}(\varepsilon) = \tfrac{1}{2\mathcal{B}}\left(c_m^{-} \pm is_m^{-}\right)$. As for the middle subspace, which is spanned by $\{\xi_m\}_{m=-N}^{N-1}$, we note that the tridiagonal requirement of the reference wave operator $H_0 - \varepsilon$ in the outer subspaces all the way to the borders of the middle subspace dictates that

$$\xi_n^{\uparrow}(x) = \phi_n^{+\uparrow}(x) = A_n e^{-y^2/2} H_{2n}(y), \tag{28a}$$

$$\xi_{-n-1}^{\uparrow}(x) = \phi_n^{-\uparrow}(x) = A_{n+\tfrac{1}{2}} e^{-y^2/2} H_{2n+1}(y), \tag{28b}$$

where $n = 0,1,..,N-1$. We also require that the subset of the basis in the middle subspace be energy independent so that numerical computations (e.g., of the potential matrix elements, Harris eigenvalues, matrix diagonalization, etc.) will be done *once* for all energies. Moreover, to make the numerical scheme very effective and simple, we maintain the tridiagonal structure of the matrix representation of the reference wave operator in the middle subspace. These requirements allow us to choose the lower components of $\{\xi_m\}_{m=-N}^{N-1}$ as follows

$$\xi_n^\downarrow(x) = 2\tau n A_n e^{-y^2/2} H_{2n-1}(y), \tag{28c}$$

$$\xi_{-n-1}^\downarrow(x) = \xi_{n+\frac{1}{2}}^\downarrow(x) = \tau(2n+1) A_{n+\frac{1}{2}} e^{-y^2/2} H_{2n}(y). \tag{28d}$$

where $\tau$ is a new dimensionless basis parameter (in addition to $\lambda$).

Now, to realize the solution of the problem (i.e., give the complete specification of the total wavefunction $\psi$) we only need to determine the $2N+2$ energy dependent quantities, $\{a_m(\varepsilon)\}_{m=-N}^{N-1}$ and $W_\pm(\varepsilon)$. Using the above results, we can write the original Dirac equation (1) in the following algebraic form

$$\langle \xi_n | (H-\varepsilon) | \psi \rangle = \sum_{m=-N}^{N-1} \left( \mathcal{J}_{n,m} + \mathcal{V}_{n,m} \right) a_m + \sum_{m=N}^{+\infty} \mathcal{J}_{n,m}^+ b_m^+ + \sum_{m=N}^{+\infty} \mathcal{J}_{n,m}^- b_m^- = 0, \tag{29}$$

where $\mathcal{J}_{n,m} = \langle \xi_n | (H_0 - \varepsilon) | \xi_m \rangle$ and $\mathcal{V}_{n,m} = \langle \xi_n | \mathcal{V} | \xi_m \rangle$. Note that $\mathcal{V}_{n,m} = 0$ in the left and right subspaces. Following the same steps as in the nonrelativistic case [1], the algebraic wave equation (29) for $n = N$ gives $a_{N-1}(\varepsilon) = b_{N-1}^+(\varepsilon)$ and for $n = -N-1$ it gives $a_{-N}(\varepsilon) = b_{N-1}^-(\varepsilon)$. It is finally reduced to the following $2N \times 2N$ matrix equation

$$\left[ \begin{pmatrix} \mathcal{J}^- & 0 \\ 0 & \mathcal{J}^+ \end{pmatrix} + \begin{pmatrix} \mathcal{V}^{--} & \mathcal{V}^{-+} \\ \mathcal{V}^{+-} & \mathcal{V}^{++} \end{pmatrix} \right] \times \begin{bmatrix} b_{N-1}^- \\ a_{-N+1} \\ \vdots \\ a_{N-2} \\ b_{N-1}^+ \end{bmatrix} = \begin{bmatrix} -\mathcal{J}_{N-1,N}^- b_N^- \\ 0 \\ \vdots \\ 0 \\ -\mathcal{J}_{N-1,N}^+ b_N^+ \end{bmatrix}. \tag{30}$$

Now, we define the finite $2N \times 2N$ matrix Green function as

$$G_{nm}(\varepsilon) = \left[ (H-\varepsilon)^{-1} \right]_{nm} = \left[ (\mathcal{J}+\mathcal{V})^{-1} \right]_{nm}, \tag{31}$$

where $\{n,m\} = -N, -N+1, \ldots, N-2, N-1$. Since the spinor basis $\{\xi_m\}_{m=-N}^{N-1}$ given by equations (28a)-(28d) is orthogonal, then the components of this matrix Green function is calculated as shown in Appendix A by using either formula (A12) or (A17). Inverting the matrix equation (30) results in the following two equations

$$-b_{N-1}^- = G_{-N,-N} \mathcal{J}_{N-1,N}^- b_N^- + G_{-N,N-1} \mathcal{J}_{N-1,N}^+ b_N^+, \tag{32a}$$

$$-b_{N-1}^+ = G_{N-1,N-1} \mathcal{J}_{N-1,N}^+ b_N^+ + G_{N-1,-N} \mathcal{J}_{N-1,N}^- b_N^-, \tag{32b}$$

which contain only two unknowns $W_\pm(\varepsilon)$ or, equivalently, $T(\varepsilon)$ and $R(\varepsilon)$. Again, in a similar manner to that of the nonrelativistic case [1], we obtain the following set of solutions for (32)

$$W_+(\varepsilon) = \left[1 - \frac{\alpha_N^+ \beta_N^+ J_+ J_- \mathcal{G}_{+-} \mathcal{G}_{-+}}{\left(1+\mathcal{G}_{++}J_+\alpha_N^+\right)\left(1+\mathcal{G}_{--}J_-\beta_N^+\right)}\right]^{-1} \left[-\rho_{N-1} \frac{1+\mathcal{G}_{++}J_+\alpha_N^-}{1+\mathcal{G}_{++}J_+\alpha_N^+}\right.$$
$$\left. + \frac{1}{\gamma_N^+} \frac{\mathcal{G}_{+-}J_-\alpha_N^+}{1+\mathcal{G}_{++}J_+\alpha_N^+}\left(\sigma_N - \sigma_{N-1}\frac{1+\mathcal{G}_{--}J_-\beta_N^-}{1+\mathcal{G}_{--}J_-\beta_N^+} + \gamma_N^+ \rho_N \frac{\mathcal{G}_{-+}J_+\beta_N^+}{1+\mathcal{G}_{--}J_-\beta_N^+}\right)\right]$$
(33a)

$$W_-(\varepsilon) = \left[1 - \frac{\alpha_N^+ \beta_N^+ J_+ J_- \mathcal{G}_{+-} \mathcal{G}_{-+}}{\left(1+\mathcal{G}_{++}J_+\alpha_N^+\right)\left(1+\mathcal{G}_{--}J_-\beta_N^+\right)}\right]^{-1} \left[\sigma_{N-1} \frac{1+\mathcal{G}_{--}J_-\beta_N^-}{1+\mathcal{G}_{--}J_-\beta_N^+}\right.$$
$$\left. - \gamma_N^+ \frac{\mathcal{G}_{-+}J_+\beta_N^+}{1+\mathcal{G}_{--}J_-\beta_N^+}\left(\rho_N - \rho_{N-1}\frac{1+\mathcal{G}_{++}J_+\alpha_N^-}{1+\mathcal{G}_{++}J_+\alpha_N^+} + \frac{\sigma_N}{\gamma_N^+} \frac{\mathcal{G}_{+-}J_-\alpha_N^+}{1+\mathcal{G}_{++}J_+\alpha_N^+}\right)\right]$$
(33b)

where we have defined $J_\pm \equiv J_{N,N-1}^\pm = -\frac{\lambda^2}{\varepsilon+M}\sqrt{N\left(N\mp\frac{1}{2}\right)}$ and the following quantities

$$\mathcal{G}_{++} = G_{N-1,N-1}, \quad \mathcal{G}_{--} = G_{-N,-N}, \quad \mathcal{G}_{+-} = G_{N-1,-N}, \quad \mathcal{G}_{-+} = G_{-N,N-1},$$ (34a)

$$\alpha_n^\pm = p_n^\pm/p_{n-1}^\pm, \quad \beta_n^\pm = q_n^\pm/q_{n-1}^\pm, \quad \gamma_n^\pm = p_n^\pm/q_n^\pm, \quad \rho_n = p_n^-/p_n^+, \quad \sigma_n = q_n^-/q_n^+.$$ (34b)

In Appendix A of [1], we have shown how to calculate the kinematical ratios defined in Eq. (34b) starting from $\{s_0^\pm, s_1^\pm, c_0^\pm, c_1^\pm\}$ and using a stable, convergent and highly accurate computational scheme. All terms in the equation pair (33) that contain $\mathcal{G}_{\pm\mp}$ indicate coupling between the odd and even scattering channels. Thus, similar to the nonrelativistic theory, if the potential matrix is even (see footnote 2) then the two channels are decoupled (i.e., $\mathcal{G}_{\pm\mp} = 0$) and we can always write

$$T(\varepsilon) = \tfrac{1}{2}\left(e^{2i\theta_+} + e^{2i\theta_-}\right) \text{ and } R(\varepsilon) = \tfrac{1}{2}\left(e^{2i\theta_+} - e^{2i\theta_-}\right),$$ (35)

where $e^{2i\theta_\pm} = W_\pm$ and

$$W_+(\varepsilon) = -\rho_{N-1}\frac{1+\mathcal{G}_{++}J_+\alpha_N^-}{1+\mathcal{G}_{++}J_+\alpha_N^+}, \quad W_-(\varepsilon) = \sigma_{N-1}\frac{1+\mathcal{G}_{--}J_-\beta_N^-}{1+\mathcal{G}_{--}J_-\beta_N^+}.$$ (36)

Now, the evaluation of the elements of the Green matrix function (shown explicitly in Appendix A) requires accurate values for the elements of the $2N\times 2N$ total Hamiltonian matrix $H_0 + \mathcal{V}$. In Appendix B, we give the exact values of the matrix elements of $H_0$ in the basis $\{\xi_m\}_{m=-N}^{N-1}$. We also show how to obtain an accurate evaluation of the matrix elements of the short-range scattering potential $\mathcal{V}_{n,m}$ by using Gauss quadrature integral approximation.

## IV. DISCUSSION AND CONCLUSION

In this paper, which is the first of a two-part sequence, we presented a formulation of the relativistic *J*-matrix method of scattering in 1+1 space-time. It is an extension of our earlier work on the nonrelativistic version of the theory [1]. Soon, we will follow with applications of the theory to various models of the relativistic scattering problem where we address interesting issues in relativistic quantum mechanics. For example, we investigate transmission resonances in the Klein energy zone where the barrier height of the vector potential exceeds twice the rest mass. We also examine the supercritical resonance near the rest mass energy in the pseudo-spin symmetric coupling mode where $S = -V$. As in the standard *J*-matrix method, it is important to note that the physical results obtained will be independent of the values of the basis parameters $\lambda$ and $\tau$ as long as these values are within the plateau of stable computations. A calculation strategy to find the plateau of stability for computational parameters is given in [19]. Moreover, the size of the plateau of stability increases with *N* and, in principle, as $N \to \infty$ the results should be independent of any choice of values for these parameters.

## APPENDIX A
## MATRIX ELEMENTS OF THE FINITE GREEN FUNCTION IN $L^2$ BASIS

The symbols in this Appendix are local and not related to those in the rest of the paper. Let $\{\psi_n\}_{n=0}^{\infty}$ be a complete $L^2$ basis in the configuration space that supports a Hermitian representation for the wave operator, $H - z$, where *H* is the Hamiltonian and *z* is a real number. The conjugate orthogonal space is spanned by $\{\bar{\psi}_n\}_{n=0}^{\infty}$, where $\langle \bar{\psi}_n | \psi_m \rangle = \langle \psi_n | \bar{\psi}_m \rangle = \delta_{nm}$ and $\sum_n |\bar{\psi}_n\rangle\langle \psi_n| = \sum_n |\psi_n\rangle\langle \bar{\psi}_n| = 1$. Thus, the matrix elements of the Green function $G_{nm}(z)$ in the basis $\{\psi_n\}$, which is formally defined by $G(z)(H - z) = 1$, is given as

$$G_{nm}(z) = \langle \bar{\psi}_n | (H - z)^{-1} | \bar{\psi}_m \rangle \tag{A1}$$

Manipulation of Green functions, which involves inverse of operators, is carried out most appropriately in an orthogonal basis $\{\chi_n\}_{n=0}^{\infty}$ in which the representation of these operators is diagonal. That is to say, we start by solving the eigenvalue problem

$$H|\chi_n\rangle = \varepsilon_n |\chi_n\rangle \tag{A2}$$

From now on, we work in a finite subspace of dimension *N*. Since the matrix representations of the relevant operators are in the basis $\{\psi_n\}$ rather than $\{\chi_n\}$, we can rewrite Eq. (A2) in the following form

$$\sum_{k=0}^{N-1} \langle \psi_m | H | \psi_k \rangle \langle \bar{\psi}_k | \chi_n \rangle = \varepsilon_n \sum_{k=0}^{N-1} \langle \psi_m | \psi_k \rangle \langle \bar{\psi}_k | \chi_n \rangle \quad ; n, m = 0, 1, .., N-1 \tag{A3}$$

where we have used the completeness property of the basis in the finite $N$ dimensional subspace,

$$\sum_k |\psi_k\rangle\langle\bar{\psi}_k| = \sum_k |\bar{\psi}_k\rangle\langle\psi_k| = I, \tag{A4}$$

and $I$ is the $N \times N$ identity matrix. In matrix notation, Eq. (A3) reads

$$\sum_{k=0}^{N-1} H_{mk} \phi_k^n = \varepsilon_n \sum_{k=0}^{N-1} \Omega_{mk} \phi_k^n \quad ; n,m = 0,1,..,N-1 \tag{A5}$$

where $\phi_k^n = \langle\bar{\psi}_k|\chi_n\rangle$ and $\Omega_{n,m}$ is the overlap matrix element $\langle\psi_n|\psi_m\rangle$. Thus, $\{\phi_k^n\}_{k=0}^{N-1}$ is the generalized eigenvector associated with the generalized eigenvalue $\varepsilon_n$. This is so because Eq. (A5) could be written as the generalized eigenvalue equation in the $\{\psi_n\}$ basis,

$$H|\phi^n\rangle = \varepsilon_n \Omega |\phi^n\rangle \tag{A6}$$

Let us define the eigenvectors matrix $\Gamma$ whose elements are $\Gamma_{n,m} \equiv \phi_n^m = \langle\bar{\psi}_n|\chi_m\rangle$. Then Eq. (A5) reads $(H\Gamma)_{m,n} = \varepsilon_n (\Omega\Gamma)_{m,n}$ which when multiplied from left by $\Gamma^\mathrm{T}$, where $\Gamma^\mathrm{T}_{n,m} = \langle\chi_n|\bar{\psi}_m\rangle$, gives

$$(\Gamma^\mathrm{T} H \Gamma)_{m,n} = \varepsilon_n (\Gamma^\mathrm{T} \Omega \Gamma)_{m,n} \quad ; n,m = 0,1,..,N-1 \tag{A7}$$

Now, the matrix $\Gamma$ simultaneously diagonalizes $H$ and $\Omega$. That is,

$$(\Gamma^\mathrm{T} H \Gamma)_{n,m} = \eta_n \delta_{n,m} \quad \text{and} \quad (\Gamma^\mathrm{T} \Omega \Gamma)_{n,m} = \sigma_n \delta_{n,m} \tag{A8}$$

Henceforth, we deduce that $\varepsilon_n = \eta_n / \sigma_n$ and equation (A1) could be written as

$$\begin{aligned} G_{nm}(z) &= \sum_{i,j,k,l=0}^{N-1} \langle\bar{\psi}_n|\chi_i\rangle\langle\chi_i|\bar{\psi}_k\rangle\langle\psi_k|(H-z)^{-1}|\psi_l\rangle\langle\bar{\psi}_l|\chi_j\rangle\langle\chi_j|\bar{\psi}_m\rangle \\ &= \sum_{i,j,k,l=0}^{N-1} \Gamma_{ni} \left\{\Gamma^\mathrm{T}_{ik}\left[(H-z\Omega)^{-1}\right]_{kl} \Gamma_{lj}\right\} \Gamma^\mathrm{T}_{jm} \end{aligned} \tag{A9}$$

Now,

$$\sum_{k,l=0}^{N-1} \Gamma^\mathrm{T}_{ik} \left[(H-z\Omega)^{-1}\right]_{kl} \Gamma_{lj} = \frac{\delta_{ij}}{\eta_i - z\sigma_i} = \frac{1}{\sigma_i} \frac{\delta_{ij}}{\varepsilon_i - z} \tag{A10}$$

Therefore, we finally obtain

$$G_{nm}(z) = \sum_{i=0}^{N-1} \frac{\Gamma_{ni}\Gamma_{mi}}{\eta_i - z\sigma_i} = \sum_{i=0}^{N-1} \frac{1}{\sigma_i} \frac{\Gamma_{ni}\Gamma_{mi}}{\varepsilon_i - z}. \tag{A11}$$

For orthogonal basis (when $\psi_n = \bar{\psi}_n$) the overlap matrix $\Omega$ is just the identity matrix $I$, hence $\sigma_i = 1$, $\eta_i = \varepsilon_i$. In this orthogonal basis, we can write

$$G_{nm}(z) = \sum_{i=0}^{N-1} \frac{\Gamma_{ni}\Gamma_{mi}}{\varepsilon_i - z}. \tag{A12}$$

**Alternative expression for $G_{nm}(z)$:**

Numerically, it is preferred to work with eigenvalues of matrices rather than their eigenvectors. Here we give an alternative formula for (A11) and (A12) where only eigenvalues are involved. Let $\overset{nm}{H}$ (and $\overset{nm}{\Omega}$) be the $(N-1) \times (N-1)$ submatrix of $H$ (and $\Omega$) obtained by deleting the $n^{\text{th}}$ row and $m^{\text{th}}$ column, respectively. The eigenvalue equation and generalized eigenvalue equation in the truncated space, which parallel equations (A2) and (A6), are

$$\overset{nm}{H}|\tilde{\chi}_k\rangle = \overset{nm}{\varepsilon_k}|\tilde{\chi}_k\rangle, \tag{A13}$$

$$\overset{nm}{H}|\tilde{\phi}^k\rangle = \overset{nm}{\varepsilon_k}\overset{nm}{\Omega}|\tilde{\phi}^k\rangle, \tag{A14}$$

where $k = 0, 1, .., N-2$. Similarly, we define the corresponding eigenvectors matrix by $\tilde{\Gamma}_{ij} \equiv \tilde{\phi}_i^j = \langle\bar{\psi}_i|\tilde{\chi}_j\rangle$ which simultaneously diagonalizes $\overset{nm}{H}$ and $\overset{nm}{\Omega}$

$$(\tilde{\Gamma}^{\mathsf{T}} \overset{nm}{H} \tilde{\Gamma})_{ij} = \overset{nm}{\eta_i} \delta_{ij} \qquad \text{and} \qquad (\tilde{\Gamma}^{\mathsf{T}} \overset{nm}{\Omega} \tilde{\Gamma})_{ij} = \overset{nm}{\sigma_i} \delta_{ij} \tag{A15}$$

and also write $\overset{nm}{\varepsilon_k} = \overset{nm}{\eta_k}\big/\overset{nm}{\sigma_k}$. Then, it could be shown that the following is an alternative but equivalent form for $G_{nm}(z)$ in a general basis

$$G_{nm}(z) = (-1)^{n+m} \frac{|\overset{nm}{\Omega}|}{|\Omega|} \left[ \frac{\prod_{j=0}^{N-2} \overset{nm}{\varepsilon_i} - z}{\prod_{j=0}^{N-1} \varepsilon_j - z} \right] = (-1)^{n+m} \frac{\prod_{i=0}^{N-2} \overset{nm}{\rho_i}}{\prod_{j=0}^{N-1} \rho_j} \left[ \frac{\prod_{i=0}^{N-2} \overset{nm}{\varepsilon_i} - z}{\prod_{j=0}^{N-1} \varepsilon_j - z} \right]. \tag{A16}$$

where $\{\rho_n\}_{n=0}^{N-1}$ and $\{\overset{nm}{\rho_k}\}_{k=0}^{N-2}$ are the eigenvalues of the overlap matrices $\Omega$ and $\overset{nm}{\Omega}$, respectively. In orthogonal basis, Eq. (A16) can be written as

$$G_{nm}(z) = (-1)^{n+m} \frac{\prod_{i=0}^{N-2} \overset{nm}{\varepsilon_i} - z}{\prod_{j=0}^{N-1} \varepsilon_j - z}. \tag{A17}$$

A by-product of formulas (A12) and (A17) is an interesting recipe to calculate the square of the elements of the normalized eigenvectors of a matrix in terms of its set of eigenvalues as follows

$$\Gamma_{nk}^2 = \frac{\prod_{i=0}^{N-2} \varepsilon_i^{nn} - \varepsilon_k}{\prod_{\substack{j=0 \\ j \neq k}}^{N-1} \varepsilon_j - \varepsilon_k}, \tag{A18}$$

which is obtained by evaluating (A12) and (A17) at $z = \varepsilon_k$ and for $n = m$.

# APPENDIX B
# MATRIX ELEMENTS OF $H_0$ AND $\mathcal{V}$ IN THE MIDDLE SUBSPACE

The top and bottom components of the finite spinor basis in the middle subspace, $\{\xi_m(x)\}_{m=-N}^{N-1}$, are given by equations (28a)-(28d). For an efficient notation, we merge these basis elements in $\{\zeta_i(x)\}_{i=0}^{2N-1}$ where $\zeta_i \equiv \xi_{i-N}$ and $i = 0,1,..,2N-1$. Using the recursion relation, differential property and orthogonality of the Hermite polynomials [14,15] we can calculate the matrix representation of the reference Hamiltonian $H_0$ in the middle subspace. As expected, this comes out to be tridiagonal having the following matrix elements (with $n,m = 0,1,..,N-1$)

$$\langle \zeta_{N+n} | H_0 | \zeta_{N+m} \rangle = \langle \xi_n | H_0 | \xi_m \rangle =$$
$$\left[ M + \tau(2\lambda - \tau M)n \right] \delta_{n,m} - \lambda\tau \left[ \sqrt{n(n-\tfrac{1}{2})} \delta_{n,m+1} + \sqrt{(n+1)(n+\tfrac{1}{2})} \delta_{n,m-1} \right] \tag{B1a}$$

$$\langle \zeta_{N-n-1} | H_0 | \zeta_{N-m-1} \rangle = \langle \xi_{-n-1} | H_0 | \xi_{-m-1} \rangle =$$
$$\left[ M + \tau(2\lambda - \tau M)(n+\tfrac{1}{2}) \right] \delta_{n,m} - \lambda\tau \left[ \sqrt{n(n+\tfrac{1}{2})} \delta_{n,m+1} + \sqrt{(n+1)(n+\tfrac{3}{2})} \delta_{n,m-1} \right] \tag{B1b}$$

Now, since $H_0$ is an even matrix operator (see footnote 2) then we also obtain $\langle \zeta_{N+n} | H_0 | \zeta_{N-m-1} \rangle = \langle \zeta_{N-n-1} | H_0 | \zeta_{N+m} \rangle = 0$. Taking the nonrelativistic limit[3] will reproduce the matrix elements of the nonrelativistic $H_0$ provided that $\tau = \lambda/M$. Now, the matrix representation of the identity (the basis overlap matrix), which is also needed in the calculation, is diagonal with the following elements

$$\langle \zeta_{N+n} | \zeta_{N+m} \rangle = \langle \xi_n | \xi_m \rangle = (1 + \tau^2 n) \delta_{n,m}, \tag{B2a}$$

---

[3] If we designate the relativistic (nonrelativistic) reference Hamiltonian as $H_0^{\text{Rel}}$ ($H_0^{\text{NRel}}$) then the non-relativistic limit ($\tfrac{\lambda}{M} \to 0$) gives: $\left( H_0^{\text{Rel}} \right)_{nm} \to M\delta_{nm} + \left( H_0^{\text{NRel}} \right)_{nm}$. However, in [1] we used atomic units where $M = 1$. Thus, for the purpose of comparing matrix elements one has to put back $M$ in [1].

$$\langle \zeta_{N-n-1} | \zeta_{N-m-1} \rangle = \langle \xi_{-n-1} | \xi_{-m-1} \rangle = \left[ 1 + \tau^2 \left( n + \tfrac{1}{2} \right) \right] \delta_{n,m}. \tag{B2b}$$

The nonrelativistic limit of these matrix elements are $\delta_{nm} + O(\lambda^2/M^2)$. On the other hand, to calculate the matrix elements of the short-range potential, $\mathcal{V} = \begin{pmatrix} V^+ & U \\ U & V^- \end{pmatrix}$ where $V^\pm(x) = V(x) \pm S(x)$, we start by defining the matrix elements of any integrable function $F(x)$ as

$$F_{n,m} = \frac{1}{\sqrt{\pi 2^{n+m} n! m!}} \int_{-\infty}^{+\infty} e^{-y^2} H_n(y) H_m(y) F(y/\lambda) dy. \tag{B3}$$

We evaluate this integral using Gauss quadrature associated with the Hermite polynomials [20-25] as follows

$$F_{nm} \approx \sum_{k=0}^{K-1} \Lambda_{nk} \Lambda_{mk} F(\eta_k/\lambda), \tag{B4}$$

where $\eta_k$ is an eigenvalue of the $K \times K$ tridiagonal quadrature matrix $\mathbb{G}$ associated with the symmetric three-term recursion relation for the normalized Hermite polynomials $\hat{H}_n(y) = \frac{1}{\sqrt{\pi 2^n n!}} H_n(y)$ (i.e., $\mathbb{G}_{nm} = \sqrt{\tfrac{n}{2}} \delta_{n,m+1} + \sqrt{\tfrac{n+1}{2}} \delta_{n,m-1}$) and $\{\Lambda_{nk}\}_{n=0}^{K-1}$ is the corresponding normalized eigenvector. The dimension $K$ of the quadrature matrix $\mathbb{G}$ should be larger than or equal to $2N$ (the size of the middle subspace). Consequently, we obtain the following elements of the potential matrix in the basis $\{\zeta_i(x)\}_{i=0}^{2N-1}$

$$\langle \zeta_{N+n} | \mathcal{V} | \zeta_{N+m} \rangle = \langle \xi_n | \mathcal{V} | \xi_m \rangle =$$
$$V^+_{2n,2m} + \tau^2 \sqrt{nm} V^-_{2n-1,2m-1} + \tau \left[ \sqrt{n} U_{2n-1,2m} + \sqrt{m} U_{2n,2m-1} \right] \tag{B5a}$$

$$\langle \zeta_{N-n-1} | \mathcal{V} | \zeta_{N-m-1} \rangle = \langle \xi_{-n-1} | \mathcal{V} | \xi_{-m-1} \rangle =$$
$$V^+_{2n+1,2m+1} + \tau^2 \sqrt{\left(n+\tfrac{1}{2}\right)\left(m+\tfrac{1}{2}\right)} V^-_{2n,2m} + \tau \left[ \sqrt{\left(n+\tfrac{1}{2}\right)} U_{2n,2m+1} + \sqrt{\left(m+\tfrac{1}{2}\right)} U_{2n+1,2m} \right] \tag{B5b}$$

$$\langle \zeta_{N+n} | \mathcal{V} | \zeta_{N-m-1} \rangle = \langle \xi_n | \mathcal{V} | \xi_{-m-1} \rangle =$$
$$V^+_{2n,2m+1} + \tau^2 \sqrt{n\left(m+\tfrac{1}{2}\right)} V^-_{2n-1,2m} + \tau \left[ \sqrt{n} U_{2n-1,2m+1} + \sqrt{m+\tfrac{1}{2}} U_{2n,2m} \right] \tag{B5c}$$

$$\langle \zeta_{N-n-1} | \mathcal{V} | \zeta_{N+m} \rangle = \langle \xi_{-n-1} | \mathcal{V} | \xi_m \rangle =$$
$$V^+_{2n+1,2m} + \tau^2 \sqrt{m\left(n+\tfrac{1}{2}\right)} V^-_{2n,2m-1} + \tau \left[ \sqrt{m} U_{2n+1,2m-1} + \sqrt{n+\tfrac{1}{2}} U_{2n,2m} \right] \tag{B5d}$$

Note that Eq. (B5a), Eq. (B5b), Eq. (B5c) and Eq. (B5d) give the respective matrix elements for $\mathcal{V}^{++}_{\text{even}}$, $\mathcal{V}^{--}_{\text{even}}$, $\mathcal{V}^{+-}_{\text{odd}}$, $\mathcal{V}^{-+}_{\text{odd}}$ in the $2N \times 2N$ matrix wave equation (30). If the potential matrix is even, which happens if $V^\pm(-x) = V^\pm(-x)$ and $U(-x) = -U(x)$, then $\mathcal{V}^{+-}_{\text{odd}} = \mathcal{V}^{-+}_{\text{odd}} = 0$. On the other hand, if $V^{\pm}(-x) = -V^{\pm}(-x)$ and $U(-x) = U(x)$ then the potential matrix is odd and $\mathcal{V}_{even}^{++} = \mathcal{V}_{even}^{--} = 0$. In the former case, the two scattering channels decouple and we end up with the scattering amplitudes given by Eq. (35).